\documentclass[twocolumn]{revtex4}
\usepackage{epsf,graphicx}
\usepackage{subfigure}
\usepackage{amsmath,amssymb,amsfonts}
\begin{document}

\author{D.H.E.~Gross, E.V.~Votyakov, and A. De Martino}

\affiliation{{$^1$}Hahn-Meitner-Institut, Bereich Theoretische
Physik, Glienickerstr. 100, 14109 Berlin (Germany)}

\title{Reply to the comment by I.Ispolatov and M.Karttunen, cond-mat/0303564}

\maketitle

The description of self-gravitating stellar systems as
equilibrized objects in the microcanonical thermodynamics applied
also to "Small" systems \cite{gross174} with sharp total energy
and angular momentum demands a spherical container centered in the
center of mass of the total system. Of course in reality there is
no such container. The container is necessary to avoid particle
loss by evaporation. The statistical equilibrium thus obtained is
a metastable equilibrium which lives over times such that the mass
loss due to evaporation is negligible, and makes sense as long as
nuclear hydrogen burning stabilizes the stars against collapse.

In order for the calculated equilibrium configuration to have
anything in common with the real free stellar system the effect of
the container on the obtained mass distribution must be minimized.
I.e., only configurations with the center of mass fixed at the
center of the container are admitted. Moreover, the density near
the container walls should be small. This was achieved in our
calculation \cite{gross187,gross190,gross191,gross195} by setting
the contributions of the spherical harmonics
$\int{b_{l=1,m}(x)x^2dx}=0$.

In the molecular-dynamics simulation by Ispolatov and Karttunen this
necessary constraint was forgotten. Therefore the spatial distribution
they show in their Figure 1 is irrelevant for reality because the
container has a non-negligible momentum and the center of mass of the
gas is not at rest. Certainly after removing this crucial restriction
it is impossible to find any realistic configuration like binary stars
rotating around their common center of mass and with a negligible
density of matter at the artificial container walls, as the one we
describe e.g. in \cite{gross187}. On the other hand, if in our setting
we release the constraint on the functions $b_{1,m}$ we get
configurations similar to those they find.

A possible way to enforce preservation of center of mass within
molecular dynamics may be to take only those initial conditions
with positive parity, where for each particle at $(r_i,p_i)$ there
is a partner starting at opposite initial point in phase space
$(-r_i,-p_i)$. Such pairs of particles would hit the container
simultaneously at opposite sides and the center of mass of the
whole system without the container would not move. This should
lead to symmetric binary configurations as the ones found by us in
\cite{gross187}.

\end{document}